\documentclass{article}
    \usepackage{spconf,amsmath,graphicx}
\usepackage{url}

\title{Remixing Music for Hearing Aids Using Ensemble of Fine-Tuned Source Separators}
\name{Matthew Daly}
\address{Johns Hopkins University \\\texttt{mdaly14@jhu.edu}}

\begin{document}
\ninept
\maketitle
\noindent\copyright\ IEEE 2024. Personal use of this material is permitted. Permission from IEEE must be obtained for all other uses, in any current or future media, including reprinting/republishing this material for advertising or promotional purposes, creating new collective works, for resale or redistribution to servers or lists, or reuse of any copyrighted component of this work in other works.
\begin{abstract}
 This paper introduces our system submission for the Cadenza ICASSP 2024 Grand Challenge, which presents the problem of remixing and enhancing music for hearing aid users.
 Our system placed first in the challenge, achieving the best average Hearing-Aid Audio Quality Index (HAAQI) score on the evaluation data set.
 We describe the system, which uses an ensemble of deep learning music source separators that are fine tuned on the challenge data.
 We demonstrate the effectiveness of our system through the challenge results and analyze the importance of different system aspects through ablation studies.
%The abstract should contain about 100 to 150words, and should be identical to the abstract text submitted electronically

\end{abstract}
\section{Introduction}
\label{sec:intro}

The Cadenza ICASSP 2024 Grand Challenge \cite{ICASSP2024-Cadenza} presents a machine learning challenge to enhance music signals for hearing aid users.
The goal of the challenge is to improve music processing for hearing aids, since hearing loss and hearing aid processing can negatively affect the music listening experience.

The task issued by the challenge is to design a system to remix a music signal for a listener according to their specific preferences and hearing loss characteristics.
The inputs to the system include an audio signal containing music and the listener-specified gains indicating their preference on how to remix the four component music tracks: vocals, drums, bass, and other (VDBO).
The ``other'' track refers to any component of the signal not belonging to the other three tracks.
The system is also provided with the listener's audiogram, which specifies their characteristic hearing loss at different frequencies.

%evaluated with haaqi
The system is evaluated by comparing the enhanced music signal with a ground-truth enhanced signal constructed using the true separate music tracks.
This comparison uses the Hearing-Aid Audio Quality Index (HAAQI) \cite{kates2015hearing}, which objectively measures the perceived audio quality of the enhanced signal given the ground-truth reference and the listener's specific hearing loss as characterized by their audiogram.

%hrtf
The Cadenza challenge provides a data set for training, a validation data set, and a test data set for the final evaluation.
The data includes songs, sets of gains for the VDBO components, and listener audiograms.
An important aspect of the data is that the audio signals are convolved with head-related transfer functions (HRTFs) to simulate a person listening to stereo speakers in a free field with different head positions.
This introduces crosstalk to the stereo channels, meaning each received stereo channel contains a combination of the separate transmitted stereo channels, depending on the head position relative to the speakers.

%baseline w/demucs
The provided baseline system for the challenge uses the Hybrid Demucs (HDemucs) model \cite{defossez2021hybrid}, trained on the MUSDB18-HQ data set \cite{MUSDB18HQ}, to separate the VDBO tracks.
The baseline remixes the music by adjusting the tracks to the desired gains and summing them together.
The baseline normalizes the signal to match the loudness of the input and then performs NAL-R hearing aid amplification \cite{byrne1986national} given the listener's audiogram.

%we focus on music source sep, rest of the system mostly the same
%describe mss
We propose a music enhancement system for the challenge that follows the baseline's basic structure, while focusing on improving the music source separation (MSS) aspect.
A variety of deep neural network architectures have demonstrated strong results on the MSS problem \cite{defossez2021hybrid,chen2023music,kim2021kuielabmdxnet,10121418}, and we use an ensemble of three different architectures, each fine tuned on the Cadenza challenge data with crosstalk.

Our proposed ensemble system placed first in the Cadenza ICASSP 2024 Grand Challenge, with the best average HAAQI score on the evaluation data set.
This paper describes the system and presents evaluation results demonstrating its effectiveness in the challenge.
We analyze the results and use ablation studies to measure the importance of different system components.

\section{System Design}

Our system follows the baseline's structure by using an MSS model to split the music into VDBO tracks, adjusting the tracks to the desired gains, summing the tracks together, performing loudness normalization, and applying NAL-R amplification.
We mainly focus on improving the MSS step by using an ensemble of three different deep neural network architectures that have achieved state-of-the-art results on the MUSDB18-HQ data set.
The model architectures are HDemucs \cite{defossez2021hybrid}, KUIELab-MDX-Net \cite{kim2021kuielabmdxnet}, and DTTNet \cite{chen2023music}.
We simply average together the VDBO signals from each of the models in the ensemble.

%describe each model and num of params
The KUIELab-MDX-Net and DTTNet architectures are both based on a Time-Frequency Convolutions with Time-Distributed Fully-connected networks UNet (TFC-TDF UNet) architecture \cite{choi_2020}.
Both of these architectures require a separate model trained for each VDBO track.
Each KUIELab-MDX-Net model contains 7.4 million parameters, and each DTTNet model contains 5 million parameters.
The HDemucs model uses a convolutional encoder-decoder architecture that operates on both frequency-domain and time-domain representations.
The HDemucs architecture requires a single model to separate the four VDBO tracks, and it contains 83.6 million parameters.

For the models in the MSS ensemble, we started with pretrained checkpoints from the official implementations, which were all trained on MUSDB18-HQ.
We then fine tuned the checkpoints on the Cadenza data, where the key difference from MUSDB18-HQ is the inclusion of HRTFs to introduce crosstalk between the stereo channels.
We trained for a maximum of 10 epochs using the same hyperparameters as the original implementations, except for reduced learning rates.
We performed early stopping by selecting model checkpoints with the highest average signal-to-distortion ratio (SDR) on the validation set.

For the DTTNet and KUIELab-MDX-Net models we selected salient segments from the training set using the source activity detection algorithm mentioned in \cite{10121418}.

Based on our experiments with the validation set, we found it useful to calculate the residual signal by subtracting the predicted vocals, drums, and bass tracks from the original music signal.
We then incorporate this residual into the final ``other'' track prediction at inference time by averaging it together with the ``other'' track predicted by the model.

We also try to reduce distortion to the signals introduced by clipping the NAL-R amplified signals by applying an audio compressor to reduce the dynamic range of the signal.
We use a rough heuristic to apply the compressor only if we detect 25 thousand samples to be clipped in one of the stereo channels.

\section{Results}

Table \ref{tab:results} displays the average HAAQI score of our system on the evaluation set, alongside the scores of the individual components of the ensemble, the HDemucs baseline system, and the systems of the next two highest scoring teams in the challenge, T003 and T011.
Our ensemble system obtained the best score out of all systems submitted to the challenge.

\begin{table}[]
    \centering
    \begin{tabular}{|l|c|}
         \hline
         \textbf{System} & \textbf{HAAQI Score} \\
         \hline
         Baseline (HDemucs) & 0.5697 \\
         Team T011 submission & 0.5857 \\
         Team T003 submission & 0.5929 \\
         Fine-tuned HDemucs & 0.6027 \\
         Fine-tuned KUIELab-MDX-Net & 0.6001 \\
         Fine-tuned DTTNet & 0.6166 \\
         \textbf{Ensemble} & \textbf{0.6317} \\
         \hline
    \end{tabular}
    \caption{HAAQI scores of systems on evaluation set.}
    \label{tab:results}
\end{table}

\subsection{Ablation Studies}

Table \ref{tab:ablations} shows the results of our ablation studies, where we tested the effect of incorporating the residual into the prediction of the ``other'' track.
We also tested the effect of applying the audio compressor.
All other aspects of the ensemble system were kept constant while removing these components for the ablation studies.
We can see both of these ablations reduced performance.

\begin{table}[]
    \centering
    \begin{tabular}{|l|c|}
         \hline
         \textbf{Ablation} & \textbf{HAAQI Score} \\
         \hline
         Ensemble (no ablation) & 0.6317 \\
         Remove use of residual & 0.6081 \\
         Remove audio compressor & 0.6053 \\
         \hline
         \end{tabular}
    \caption{HAAQI scores of ablation studies.}
    \label{tab:ablations}
\end{table}

\section{Discussion}

The results show how our system was successful in the challenge.
We can compare the fine-tuned HDemucs model system to the baseline system to see that fine tuning on the data with crosstalk leads to improved performance.
We can also see the benefit of averaging together results in the ensemble.
This suggests the different model architectures are able to complement each other on the separation task. %possibly add something here

The ablation studies also show that the incorporation of the residual signal and the application of the audio compressor are important factors in the system's success.

\section{Conclusions}
We introduced a successful system for the Cadenza ICASSP 2024 Challenge that uses an ensemble of source separators to remix music and enhance it for hearing aid users.
Our results demonstrate the effectiveness of using an ensemble, fine tuning pretrained models, incorporating the residual signal, and using an audio compressor.
These components of our system led to it achieving the highest score and placing first in the challenge.

\bibliographystyle{IEEEtran}
\bibliography{refs}

\end{document}